\documentstyle[aps,preprint]{revtex}
\begin{document}
\draft
\tightenlines
 
\title{Interface Scaling in the Contact Process}
\author{Ronald Dickman$^{1,*}$ and Miguel A. Mu\~noz$^{2,\dagger}$}
\address{
$^1$Departamento de F\'{\i}sica, ICEx,
Universidade Federal de Minas Gerais,\\
Caixa Postal 702, 30161-970
Belo Horizonte - MG, Brasil\\
$^2$ Institute {\it Carlos I} for Theoretical and Computational Physics\\
and Departamento de Electromagnetismo y F{\'\i}sica de la Materia,\\
Universidad de Granada, 
18071 Granada, Spain.\\}
\date{\today}

\maketitle
\begin{abstract}
Scaling properties of an interface representation of the critical 
contact process are studied in dimensions 1 - 3.  Simulations confirm 
the scaling relation $\beta_W = 1 - \theta$ between the 
interface-width growth exponent 
$\beta_W$ and the exponent $\theta$ governing the decay of the order 
parameter. A scaling property of the height distribution, which serves as
the basis for this relation, is also verified.  The height-height correlation 
function shows clear signs of anomalous scaling, in accord with L\'opez' 
analysis [Phys. Rev. Lett. {\bf 83}, 4594 (1999)], but no evidence of 
multiscaling.
\end{abstract}


\section{Introduction}

Scaling and criticality in nonequilibrium systems continue to be of great interest
in statistical physics.  Among the various classes of systems that have been subject
to intensive study are models of growing interfaces \cite{spohn,hhz,barabasi,krug},
and absorbing-state phase transitions, typified by directed percolation (DP)
\cite{kinzel,torre,harris,revs}.  Absorbing-state phase transitions have been linked
to self-organized criticality (SOC) \cite{vz,dvz,paths}, as have driven interface models
\cite{pacz,midd,lau,ala}.  The latter connection is established by defining
a height variable $h_i(t)$ at each site of a sandpile model; at each time interval, the
height increases by one unit at each active (toppling) site.  It turns out that the interface
of the Bak-Tang-Wiesenfeld model \cite{btw}
is described by an Edwards-Wilkinson equation with columnar noise \cite{lau,ala}.
In view of the connections between absorbing-state phase transitions, SOC, and
surface growth,
it is of interest to study the dynamics of the interface representation of a simple 
model in the DP class.  Precise results on the scaling
properties of the DP interface representation should prove useful when trying to
assign interface models (or height representations of other models, such as
sandpiles) to universality classes.

In this paper we examine the dynamics of the contact process in dimensions 1 - 3, 
studying the interface width and the height-height correlation
function, as well as the height probability distribution.  In Sec. II we define the model
and then present a brief discussion of the associated continuum equation, and of a 
scaling theory.  Numerical results are presented in Sec. III. Sec. IV contains a brief
summary.

\section{Model}

The contact process (CP) is a simple particle system (lattice Markov process)
exhibiting a phase transition to an absorbing
(frozen) state at a critical value of the creation rate \cite{liggett}.  This model belongs to the
universality class of directed percolation \cite{kinzel} and Reggeon field theory \cite{rft},
and is pertinent to models of epidemics \cite{harris},  catalysis \cite{catal}, 
and damage spreading \cite{damage}, among many others.
In the CP each site of the hypercubic lattice ${\cal Z}^d$ 
is either vacant or occupied by a particle.
Particles are created at vacant sites at rate $\lambda n /2d$, where $n$ is
the number of occupied nearest-neighbors, and
are annihilated at unit rate, independent of the surrounding
configuration. The order parameter is the particle
density $\rho$; the vacuum state, $\rho = 0 $, is absorbing.
As $\lambda$ is increased beyond $\lambda_c$,
there is a continuous phase transition from the vacuum 
to an active state; for $\lambda > \lambda_c$, 
$\rho \sim (\lambda - \lambda_c)^{\beta}$ in the stationary state.
In one dimension, $\lambda_c \simeq 3.297848$.

There are a number of ways (equivalent as regards scaling behavior), 
of implementing the CP in a simulation algorithm;
this work follows the widely used practice of maintaining a list
of all occupied sites.  In this study the initial condition is always that of all sites
occupied.  Subsequent events involve selecting (at random) an occupied
site {\bf x} from the $N_p$ sites on the list, selecting a process: creation with probability
$p = \lambda/(1+\lambda)$, annihilation with probability $1-p$, and, in the
case of creation, selecting one of the $2d$ nearest neighbors, {\bf y}, 
of {\bf x}.  (The creation attempt succeeds if {\bf y} is vacant).
The time increment $\Delta t$ associated with an event is $1/N_p$, where $N_p$
is the number of occupied sites immediately prior to the event.
A trial ends when all the particles have vanished, or at the
first event with time $\geq t_m$, a preset maximum time. 

The important scaling laws pertinent to the critical contact process on a lattice of
$L^d$ sites, starting with all sites occupied are: (1) the mean survival time
$\tau \propto L^{\nu_{||}/\nu_\perp}$, and (2) the average particle density decays
as a power law, $\rho (t) \propto t^{-\theta}$ for $1 < t < \tau$.  
(In practice, the power law is found already for $t \approx 2$.)
In one dimension,
$\nu_{||}/\nu_\perp \simeq 1.5808$ and $\theta \simeq 0.1595$ \cite{iwancp}.

Occupied sites represent activity, which spreads from site to site.  (The absence of
any activity corresponds to the absorbing state.)  In a growing surface or driven interface,
activity corresponds to the motion of the interface.  We therefore define the {\it height}
$h_i(t)$ at site $i$ as the amount of time (up to time $t$) that site $i$ has been
occupied.  In our numerical studies we use a real-valued height (recall that time is
not restricted to integer values in our implementation).  By keeping a record of the
last time $t_i$ at which the state of site $i$ changed, we are able to evaluate
$h_i(t)$ at any moment in the simulation.  (While the results reported here are for
real $h$, we find the same scaling properties for integer $h$.)
The surface $h_i(t)$ may be thought of as a driven interface.
Since the critical contact process must eventually enter the absorbing state, the
interface, in this analogy, will eventually be pinned.

The large-scale properties of the CP and related models such as DP can be described 
via a field theory (so-called Reggeon field theory) framed in terms of a coarse-grained 
density $\rho ({\bf x},t) \geq 0$ \cite{rft}.  Retaining only relevant terms, the 
stochastic pde for $\rho ({\bf x},t)$ reads
\begin{equation}
\frac{ \partial \rho}{\partial t} = \nabla^2 \rho -a \rho - b\rho^2 + \eta(x,t) \;.
\label{cpft}
\end{equation}

\noindent Here $\eta ({\bf x},t)$ is a Gaussian noise 
with zero mean and autocorrelation

\begin{equation}
\overline{ \eta({\bf x},t) \eta({\bf x}',t') } = \Gamma \rho({\bf x},t) 
\delta^d({\bf x-x}') \delta(t-t') \;.
\label{noisecp}
\end{equation}
In our continuum description, the height is given by

\begin{equation}
h({\bf x},t) = \int_0^t dt' \rho({\bf x}, t') \;.
\label{htcont}
\end{equation}
Integrating Eq. (\ref{cpft}) from time zero to time $t$, we obtain

\begin{equation}
\frac{ \partial h}{\partial t} = \nabla^2 h -a h - 
b\int_0^t \left(\frac{\partial h}{\partial t'} \right)^2 dt' + \zeta({\bf x},t) \;,
\label{hft}
\end{equation}
with the noise autocorrelation

\begin{equation}
\overline{ \zeta({\bf x},t) \zeta({\bf x}',t') } = \Gamma  \delta^d({\bf x-x}') 
h({\bf x},t_<) \;,
\label{hnoise}
\end{equation}
where $t_< \equiv \min(t,t')$.
Thus the equation governing the height includes a nonlinear memory term
and a noise with nonvanishing correlations between different times.  This equation
does not seem to shed much light on the scaling properties of the interface.
Rather, the relation between $h$ and the density in the contact process,
Eq. (\ref{htcont}), yields some properties that are not
immediately obvious from Eq. (\ref{hft}), for example, that $\partial h/\partial t \geq 0$,
and that the nonlinear term is relevant for $d < d_c = 4$.
In any event, Eq. (\ref{hft}) does serve to point up some differences between
the CP interface and conventional surface-growth models such 
as the Edwards-Wilkinson \cite{ew}
or Kardar-Parisi-Zhang equations \cite{kpz}.  
First, the linear ``drive" term ($-a h({\bf x},t)$ with $a<0$
in the active regime) is proportional to the local height and so cannot be transformed
away.  Second, in the active state, the moduli of the last three terms on the r.h.s. of 
Eq. (\ref{hft}) grow without limit, suggesting that $\nabla^2 h$ does as well,
so that the width of the active phase never saturates.

We turn now to a simple scaling analysis of the interface.  At its basis lies
a scaling hypothesis for the probability density $p(h;t)$ of the height $h$ (at any
lattice site) at time $t$: the time-dependence of this density
enters only through the {\it mean height} $\overline{h}(t)$.  The conjectured scaling
property is

\begin{equation}
p(h;t) = \frac{1}{\overline{h}(t)} {\cal P} (h/\overline{h}(t)) \;,
\label{schyp}
\end{equation}
where the scaling function ${\cal P} \geq 0$ with $\int {\cal P}(u) du = 1$, 
and the prefactor guarantees normalization.  It follows that the moments of
$h$ all scale with the mean height, 
$\overline{h^n} = \overline{u^n} [\overline{h}(t)]^n$ with $\overline{u^n}$ the
$n$-th moment ${\cal P}$.  In particular, the mean-square width

\begin{equation}
W^2 (t,L) = \mbox{var}(h) \sim [\overline{h}(t)]^2 \;,
\label{scw2}
\end{equation}
if Eq. (\ref{schyp}) holds.  On the other hand, we have that for times $t < \tau$
in the critical contact process,

\begin{equation}
\overline{h}(t) = \int_0^t dt' \rho(t') \sim t^{1-\theta} \;,
\end{equation}
which immediately implies that $W^2 \sim t^{2 \beta_W} $ with 
$\beta_W = 1 - \theta$.  

In surface growth studies the crossover time is expected to
scale as $t_{\times} \sim L^z$; for a process in the DP universality class,
we may write $z = \nu_{||}/\nu_\perp$ (clearly $t_{\times}$ and $\tau$
should scale with the same exponent).  Then the roughness exponent $\alpha$,
defined via $W^2 (t,L) = W_{sat}^2 (L) \sim L^{2 \alpha} $ for $t \gg t_{\times}$, is given by
the scaling relation $\alpha = \beta_W z = (1 - \theta) \nu_{||}/\nu_\perp$.
It is perhaps worth stressing that the expressions relating $\alpha$
and $\beta_W$ to DP exponents depend on the validity of the scaling
hypothesis, Eq. (\ref{schyp}), which should be tested.  Inserting the known
DP exponent values, scaling theory
yields $\alpha \simeq 1.3287$, 0.97, 0.51, and zero for dimensions
1, 2, 3, and 4, respectively.  In other words, the interface 
associated with the critical contact process is super-rough in
one dimension, and asymptotically flat in $d \geq d_c = 4$, where $\theta = 1$. 

The family of height-difference correlation functions

\begin{equation}
G_q (r,t) \equiv \overline {|h(x,t) - h(x+r,t)|^q}
 \;,
\label{corrfct}
\end{equation}
are also much studied in surface growth processes.  Starting with a flat interface
at $t=0$, we expect power-law growth, $G_q \sim r^{q \alpha_q}$,
for $r < \xi(t) \sim t^{1/z}$, the time-dependent correlation length.  
If $\alpha_q$ depends on $q$ the interface is said to
be multi-affine.  For $r > \xi$, $G_q$ will saturate; in particular,
$G_2$ should approach $W^2(t,L) \sim t^{2\beta_W} \sim \xi^{2\alpha} $
for $r \approx \xi$ and $t \ll t_{\times}$.  Since $\xi $ is the only length-scale relevant to
correlations at short times (i.e., for $t < t_{\times}$, so that the system
size $L$ does not come into play), it is reasonable to expect the scaling
form

\begin{equation}
G_2 (r,t) = \xi^{2 \alpha} {\cal G} (r/\xi) \;,
\label{g2sc}
\end{equation}
where the scaling function ${\cal G}(x) \sim x^{2 \alpha_2} $ for small $x$
and is constant for large $x$.  The case $\alpha_2 = \alpha$ is referred to
as ``conventional" scaling while $\alpha_2 < \alpha$ is characterized
as ``anomalous" scaling \cite{krug,jmlopez}.

\section{Simulation Results}

We determined the square width $W^2 (t,L) $ for rings of $L = 500$,
1000, 2000, and 5000 sites, in samples of 2000, 1000, 1000, and 400 trials,
respectively.  The maximum time ranged from about $1.6 \times 10^5$
for $L=500$ to $8.8 \times 10^6$ for $L=5000$.  
All simulations were performed at the critical point,
$\lambda_c \simeq 3.297848$. In the contact process,
interfacial properties can be studied over the full sample, or over a
sample restricted to those trials that survive to time $t$.  When a trial enters the 
absorbing state,
$W^2$ naturally remains fixed for all subsequent times, and since
all trials do eventually reach the absorbing state in the critical contact process,
the interface width saturates for large $t$.  If we restrict the sample to
surviving trials, however, there is no saturation.  It is important to note
that the same scaling laws apply in either case.  (The power-law
growth regime, for example, corresponds to times $< \tau$, for which
all trials still survive.)

Fig. 1 shows a series of snapshots of the interface in a single trial with $L=200$,
at intervals of 5000 time units.  The progressive roughening of the interface,
without evidence of saturation, is apparent.
Fig. 2 is a scaling plot of the square width, averaged over all trials, i.e.,
$W^2 (L,t)/L^{2\alpha}$ versus
$\tilde{t} \equiv t/L^z$, using the exponents $\alpha = 1.328$ and $z = 1.5808$ expected
for DP in 1+1 dimensions.  There is a near-perfect
data collapse for $\tilde{t} > 10^{-3}$.  The power-law portion of the graph 
($10^{-3} \leq \tilde{t} \leq 0.05$) yields the
growth exponent $\beta_W = 0.839(1)$, in good agreement with the value $1-\theta = 0.8405$
expected on the basis of the scaling argument.
(Figures in parentheses denote statistical uncertainties.)
Analysis of the saturation width yields $W^2_{sat} (L) \sim L^{2\alpha}$ with
$\alpha = 1.325(15)$.

In Fig. 3 we test the scaling assumption for the height probability distribution $p(h)$
by plotting $t^{0.84} p(h)$ versus $h/t^{0.84}$.  (Recall that $\overline h$ is
expected to grow $\propto t^{1-\theta} = t^{0.84}$.)  In this
system of $L=1000$ sites, there is a near-perfect data collapse, in accord with Eq. (\ref{schyp}), 
for times between 500 and $10^4$. For $t = 2 \times 10^4$ we begin to note a
departure from scaling, which becomes more pronounced at later times.
Note that $t=2 \times 10^4$ corresponds to $\ln \tilde{t} = -1$, which is where
$W^2$ begins to depart noticeably from a power law in Fig. 2.
Analysis of  $p(h)$ for $L=5000$ yields similar results.  The form of the scaling function
for a given reduced time $\tilde{t}$ is independent of the system size, as shown in Fig. 4,
which compares height probability distributions for systems of 1000 and 5000 sites
at times corresponding to the same $\tilde{t}$.

The height-height correlation function $G_2$ in a system of 
1000 sites is shown in Fig. 5, which is a double-logarithmic plot of 
$G^*_2 \equiv G_2/t^{2\beta_W} $ versus $r* \equiv r/t^{1/z}$.  There
is a perfect data collapse for $t$ = 500,...,$10^4$ (reduced times 
$10^{-2} \leq \tilde{t} \leq 0.2$), with a power-law portion $G_2 \sim r^{2\alpha_2}$
with $\alpha_2 \simeq 0.625$.  At later times $G^*_2$ does not collapse, principally because
the square width has begun to saturate (it no longer grows $\propto t^{2\beta_W} $).
The value of $\alpha_2$ is insensitive to system size.  For $L=5000$ we obtain
$\alpha_2 = 0.623(2)$ for $t=2 \times 10^5$, and 0.644 for $t = 5 \times 10^5$.
Thus we may, with a high degree of confidence, adopt the estimate
$\alpha_2 = 0.63(3)$, clearly much smaller than the roughness
exponent $\alpha = 1.33$ found in the analysis of the square width,
indicating that this one-dimensional system exhibits anomalous scaling.

Recently, L\'opez argued that anomalous surface roughening is
associated with a diverging height gradient \cite{jmlopez}.  In view of the growing spikes
evident in the profiles shown in Fig. 1, this association seems very probable
in the present instance.  Indeed, the mean-square height gradient

\begin{equation}
s(t) = \overline{(\nabla h)^2} \;.
\label{sqgrad}
\end{equation}
diverges as a power law (see Fig. 6).
We find
$s(t) \sim t^{2\kappa}$, with $\kappa = 0.439(1)$ for $L=1000$,
$0.4335(4)$ for $L=2000$, and 0.4337(4) for $L=5000$; we adopt
$\kappa = 0.4336(4)$ as our best estimate.  From L\'opez'
analysis, one expects

\begin{equation}
\alpha_q = \alpha - z\kappa.
\label{lopezsc}
\end{equation}
Inserting our result for $\kappa$ and the DP values 
$\alpha = 1.32867(14)$ and $z=1.5808(1)$
in the r.h.s., we obtain $\alpha_q = 0.643(1)$,
in good agreement with the result found from analysis of the correlation
function.

We also studied the generalized height-height correlation function
$G_q$, Eq. (\ref{corrfct}), for $q$ = 1/2, 1, 3/2, 2 and 3, in a system with $L=1000$,
at $t=10^4$, and found that the functions $[G_q]^{1/q}$ are identical, and
give $\alpha_q \simeq 0.62$.  We may
therefore conclude that the DP interface is self-affine not multi-affine.

In two dimensions we studied systems of up to 256$\times$256 sites
at the critical point, $\lambda_c \simeq 1.6488$.
The curves for $W^2$ again show a good collapse (see Fig. 7), and
the derived exponents are in very good agreement with the expected
values.  Specifically, we find $\alpha= 0.97(1)$, $\beta_W = 0.550(5) $,
and $z= 1.765(10)$.  
Scaling relations combined with known DP exponent
values yield $\alpha = 0.970(5)$ and $\beta = 0.549(2)$, while current
best estimates give $z = \nu_{||}/\nu_{\perp} = 1.766(2)$ for DP in
2+1 dimensions \cite{gz,voigt,rew,tnote}.

For smaller system sizes we observe a transient in $W^2$ and all
other measured quantities, which does not appear in one dimension.
Anomalous roughening is also seen in two and three dimensions.
Fig. 8 shows $s(t)$ growing $\sim t^{2 \kappa}$ with $\kappa = 0.33(1)$.
The height-height correlation function exhibits a good collapse, as shown in 
Fig. 9; the initial power-law growth yields $\alpha_2 = 0.385(5)$;
thus the scaling relation Eq. (\ref{lopezsc}) is well satisfied:
$\alpha_2 - \alpha + z \kappa = 0.00(3)$.

We determined the interface exponents in three dimensions
($\lambda_c \simeq 1.3169$, system size $50^3$ sites, maximum time
10$^4$),
though to somewhat lower precision, owing to the larger computational
demand.  The scaling relation yields
$\beta_W = 1 - \delta = 0.274(1)$, while our simulation results
for $W^2$ give $\beta_W = 0.27(1)$.
In three dimensions we find $\alpha=0.51(1)$, $z=1.90(5)$, and $\kappa = 0.22(1)$.
Together with Eq. (\ref{lopezsc}), these yield $\alpha_2 = 0.09(2)$.
Our results for critical exponents are collected in Table \ref{tab1}.

\section{Summary}

Defining an interface representation for the contact process by analogy
with similar representations for sandpile models, we verified the expected
scaling relation $\beta_W = 1 - \theta$ in dimensions 1 - 3, and the scaling property 
of the height probability distribution in one dimension.  The local roughness 
exponent, $\alpha_2$, is smaller than the global value, $\alpha$, indicating anomalous 
surface growth.  This anomalous scaling is attended
by a diverging local slope, $s(t) = \overline{(\nabla h)^2} \sim t^{2\kappa}$.  Our
results for $\kappa$ are consistent with the scaling relation Eq. (\ref{lopezsc})
derived by L\'opez. There is, on the other hand, no
evidence of multi-affinity in this process.

An interesting point is that the process continues to
exhibit anomalous scaling for $d = 2$ and 3, even though $\alpha < 1$ in 
these cases.  While it was pointed out some time ago that $\alpha > 1$ implies
anomalous growth \cite{lesctang}, the latter appears to be an intrinsic feature
of the contact process (and, by extension, of other models in the DP
universality class) below $d_c$.

Finally, we note that the anomalous roughening analysis introduces 
two new critical exponents,
$\alpha_2$ and $\kappa$, and only one new scaling relation between them.
We therefore have a new independent exponent, $\kappa$ for instance, 
that is not related to 
the standard DP exponents in any way. A very interesting theoretical task
is that of computing $\kappa$ in an epsilon expansion around the upper critical
dimension $d_c=4$.  Our guess is that this new anomalous exponent
is related to the renormalization of a composite operator not consider so far
in the analysis of the Reggeon field theory,
but this issue is beyond the scope of this paper, and
will be studied elsewhere. 
\vspace{2em}

\noindent {\bf Acknowledgements}
\vspace{1em}

\noindent We are grateful to A. Allbens, J. Krug, J. M. L\'opez and J. G. Moreira
for helpful comments.  This work was supported in part by CNPq, by the
European Network Contract ERBFMRXCT980183, and by the Ministerio de Educaci\'on,
under project DGESEIC, PB97-0842.       
CAPES is acknowledged for support of computational facilities.
\vspace{1em}

\noindent \vspace{1em}
 
\noindent $^*${\small electronic address: dickman@fisica.ufmg.br } \\
\noindent $^\dagger${\small electronic address: mamunoz@onsager.ugr.es } \\

\newpage

\begin{table}
\begin{center}
\begin{tabular}{|c|c|c|c|c|c|}
Dimension   & $\alpha$ &  $\alpha_2$ & $\beta_W$ & $ \kappa $ &  $z$    \\
\hline
\hline
 $d=1$      &  1.33(1)   &    0.63(3)      &   0.839(1)  &  0.4336(4)    &  1.58(1)  \\        
 \hline
 $d=2$      &  0.97(1)   &    0.385(5)      &   0.550(5)  &  0.33(1)    &  1.765(10)  \\ 
 \hline    
 $d=3$      &  0.51(1)   &    0.09(2)      &   0.27(1)  &  0.22(1)    &  1.90(5)  \\
\hline
 $d=4$     &   0       &    0       &    0     &   0      &   2    \\
\end{tabular}
\end{center}
\label{tab1}
\noindent{Summary of interface-growth critical exponents for the contact process obtained
in our simulations ($d=1$-3).  Figures in parentheses denote uncertainties.
}  
\end{table}

\newpage

\noindent FIGURE CAPTIONS
\vspace{1em}

\noindent FIG. 1.  Interface of the critical contact process in a system of 200
sites, shown at intervals of 5000 time units.
\vspace{1em}

\noindent FIG. 2. Scaled mean-square width versus reduced time for system
sizes $L=500$, 1000, 2000, and 5000.
\vspace{1em}

\noindent FIG. 3. Scaling plot of the height probability distribution 
(unnormalized) for $L=1000$
at (from top to bottom on left-hand side) times 
500, 1000, 2000, 5000, $10^4$, $2 \times 10^4$, $5 \times 10^4$, and $10^5$.
\vspace{1em}

\noindent FIG. 4. Scaling plot of the height probability distributions 
(unnormalized) for 
$L=1000$ (dashed lines) and $L=5000$ (solid lines) at reduced times
(from top to bottom on left-hand side) $\tilde{t} = 0.142$, 0.71, and 1.42.
\vspace{1em}

\noindent FIG. 5. Scaled height-height correlation function versus
$r/t^{1/z}$ for $L=1000$.  The topmost curve comprises collapsed
data for times 500, 1000, 2000, 5000, and $10^4$; below it lie results for
$t = 2 \times 10^4$ (solid curve), $5 \times 10^4$ (dotted curve), 
and $10^5$ (dashed curve).
\vspace{1em}

\noindent FIG. 6. Growth of the mean-square gradient 
$s(t) = \overline{(\nabla h)^2} $ in systems with $L=1000$ (+),
2000 (dashed line), and 5000 (solid line).
\vspace{1em}

\noindent FIG. 7. Scaled mean-square width versus reduced time in
two dimensions; system sizes $L=32$, 64, 128, and 256.
\vspace{1em}

\noindent FIG. 8. Growth of the mean-square gradient 
$s(t) = \overline{(\nabla h)^2} $ in the two-dimensional system with $L=256$.
\vspace{1em}

\noindent FIG. 9. Scaled height-height correlation function versus
$r/t^{1/z}$ for $L=256$ in two dimensions.  The curves (bottom to top)
correspond to $r = 2$, 4, 8, and 16.


\begin{thebibliography}{99}

\bibitem{spohn}
	J. Krug and H. Spohn, in
	{\it Solids Far From Equilibrium: Growth, Morphology, and Defects},
	G. Godr\`eche, ed. (Cambridge University Press, Cambridge, 1990).

\bibitem{hhz}
	T. Halpin-Healy and Y.-C. Zhang,
	Phys. Rep. {\bf 254}, 215 (1995).

\bibitem{barabasi}
        A. -L. Barab\'asi and H. E. Stanley,
        {\it Fractal Concepts in Surface Growth},
        (Cambridge University Press, Cambridge, 1995).

\bibitem{krug}
	J. Krug,
	Adv. Phys. {\bf 46}, 139 (1997).

\bibitem{kinzel}
       W. Kinzel,
       Z. Phys. B{\bf 58}, 229 (1985).
  
\bibitem{torre}
         P. Grassberger and A. de la Torre,
         Ann. Phys. (N.Y.) {\bf 122}, 373 (1979).

\bibitem{harris} 
	T.E. Harris, 
	Ann. Prob. {\bf 2}, 969 (1974).

\bibitem{revs}
	R. Dickman,
	in {\em Nonequilibrium Statistical Mechanics in One Dimension}
	V. Privman, Ed. (Cambridge University Press, Cambridge 1996);
            G. Grinstein and M. A. Mu\~noz, in 
	{\it Fourth Granada Lectures in Computational Physics}, 
	Ed. P. Garrido and J. Marro, 
	Lecture Notes in Physics, {\bf 493}, 223 
	(Springer-Verlag, Berlin, 1997).
	J. Marro and R. Dickman,
	{\em Nonequilibrium Phase Transitions in Lattice Models}
	(Cambridge University Press, Cambridge, 1999).
	
\bibitem{vz}
	A. Vespignani and S. Zapperi,
	Phys. Rev. Lett. {\bf 78}, 4793 (1997);
	Phys. Rev. E {\bf 57}, 6345  (1998).

\bibitem{dvz}
	R. Dickman, A. Vespignani and S. Zapperi,
	Phys. Rev. E {\bf 57}, 5095 (1998).

\bibitem{paths}
	R. Dickman, M. A. Mu\~noz, A. Vespignani and S. Zapperi,
	e-print: cond-mat/9910454.

\bibitem{pacz} 
	M. Paczuski and S. Boettcher,  
	Phys. Rev. Lett. {\bf 77}, 111 (1996).

\bibitem{midd} 
	O. Narayan and A. A. Middleton, 
	Phys. Rev. B {\bf 49} 244 (1994).

\bibitem{lau}
	K. B. Lauritsen and M. Alava, 
	e-print: cond-mat/9903346.

\bibitem{ala}
	M. Alava and K. B. Lauritsen,
	cond-mat/0002406.

\bibitem{btw}
	P. Bak, C. Tang and K. Wiesenfeld,
	Phys. Rev. Lett. {\bf 59}, 381 (1987);
	Phys. Rev. A {\bf 38}, 364 (1988).

\bibitem{liggett}
	T. M. Liggett,
	{\it Interacting Particle Systems},
	Springer-Verlag, (New York, 1985).
 
\bibitem{rft}
        P. Grassberger,
        Z. Phys. B {\bf 47}, 365 (1982);
        H.K. Janssen,
        Z. Phys. B {\bf 42}, 151 (1981).
                                                                            
\bibitem{catal} 
        R. M. Ziff, E. Gulari, and Y. Barshad,
        Phys. Rev. Lett. {\bf 56}, 2553 (1986);
        G. Grinstein, D.-W. Lai, and D. A. Browne,
        Phys. Rev. A{\bf 40}, 4820 (1989). 
 
\bibitem{damage}
        P. Grassberger,
        J. Stat. Phys. {\bf 79}, 13 (1995).

\bibitem{iwancp}
        I. Jensen,
        J. Phys. A {\bf 29}, 7013 (1996).

\bibitem{ew} 
        S. F. Edwards and D. R. Wilkinson,
        Proc. Roy. Soc. London A {\bf 381}, 17 (1982).

\bibitem{kpz} 
        M. Kardar, G. Parisi, and Y.-C. Zhang,
        Phys. Rev. Lett. {\bf 56}, 889 (1986).

\bibitem{jmlopez} 
        J. M. L\'opez,
        Phys. Rev. Lett. {\bf 83}, 4594 (1999).

\bibitem{gz}
         P. Grassberger and Y. Zhang, 
         Physica A{\bf 224}, 169 (1996). 

\bibitem{voigt}
      C. A. Voigt and R. M. Ziff,
      Phys. Rev. E{\bf 56}, R6241 (1997).

\bibitem{rew}
      R. Dickman,
      Phys. Rev. E{\bf 60}, R2441 (1999).

\bibitem{tnote}
	Note that we use the well known scaling relation $\theta = \delta$
	in this analysis.

\bibitem{lesctang}
            H. Leschhorn and L.-H. Tang,
            Phys. Rev. Lett {\bf 70}, 2973 (1993).

\end{thebibliography}
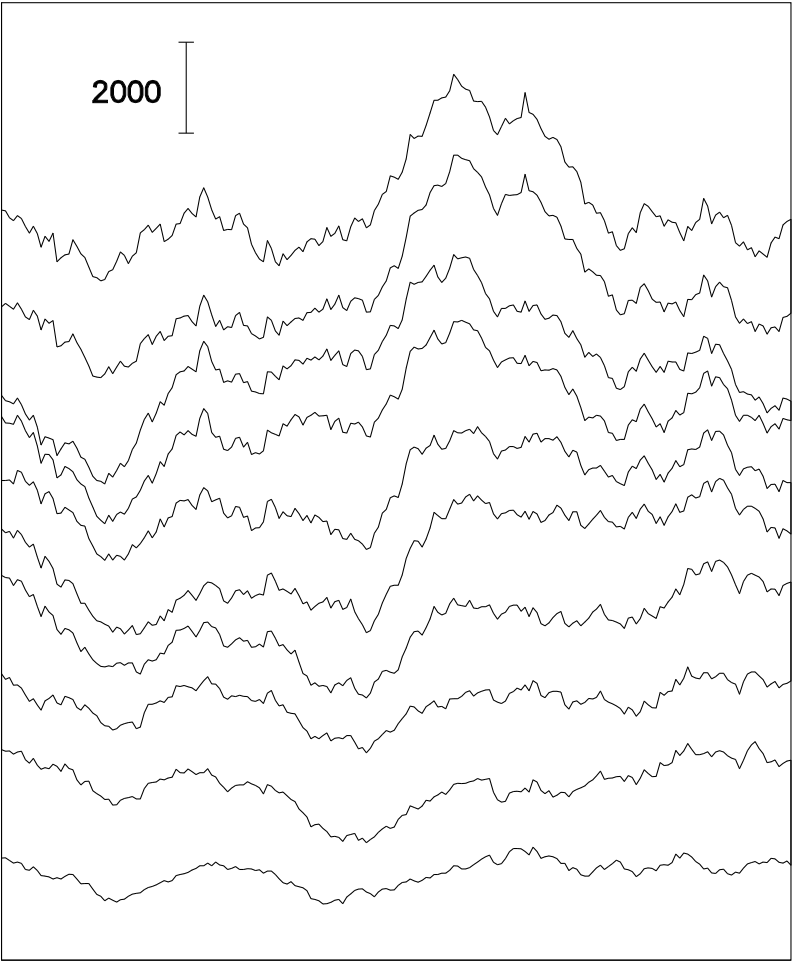
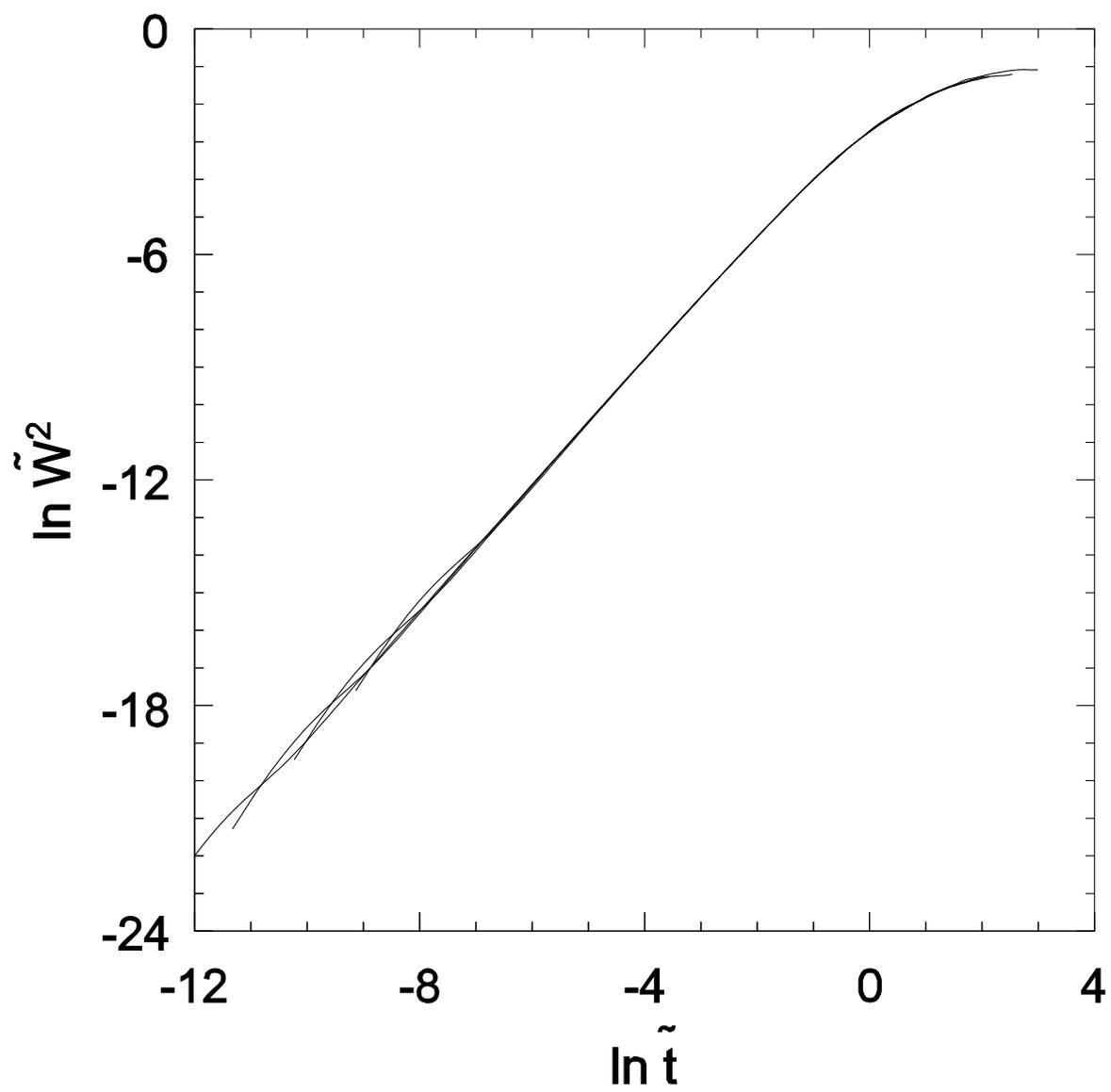
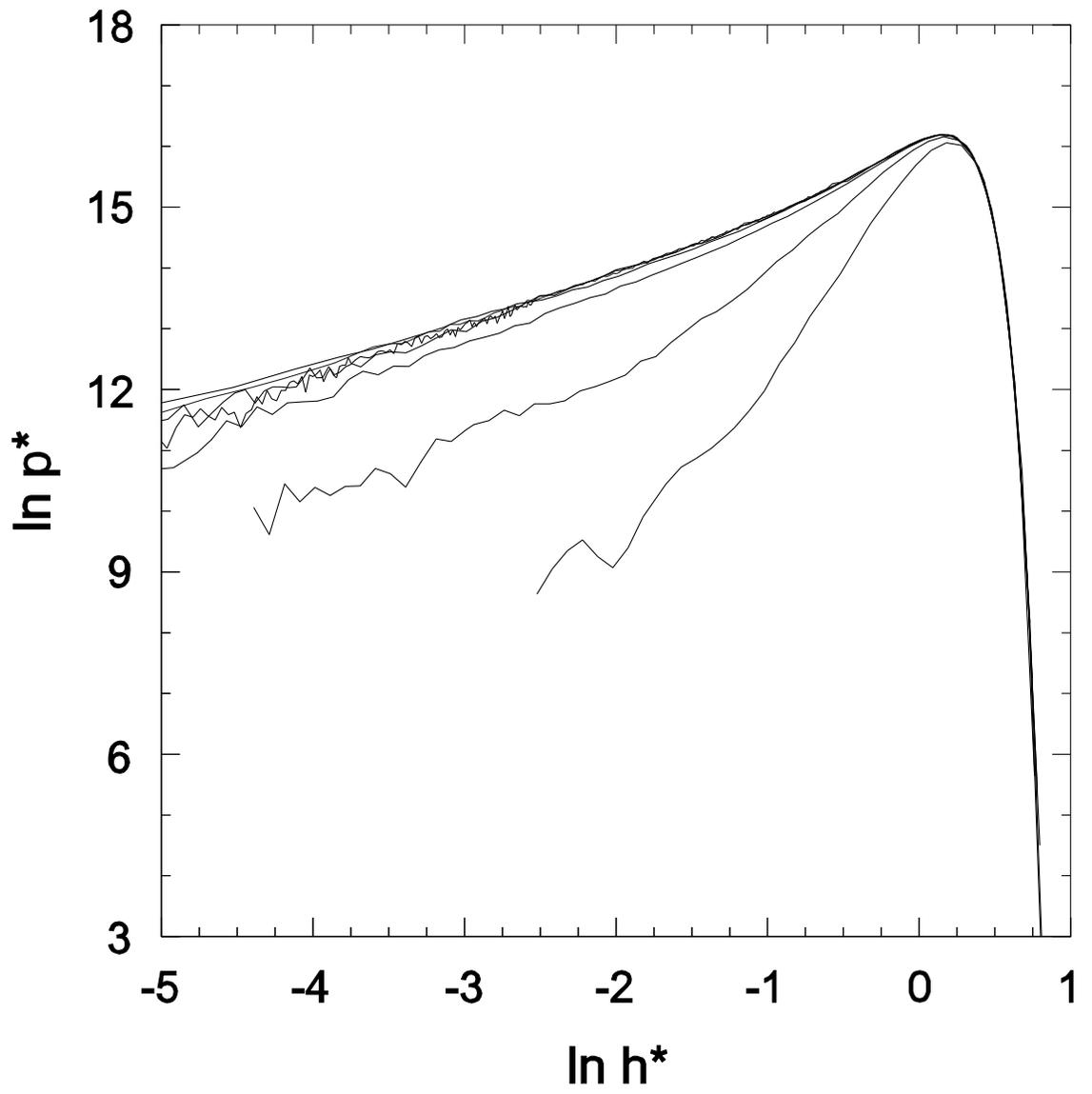
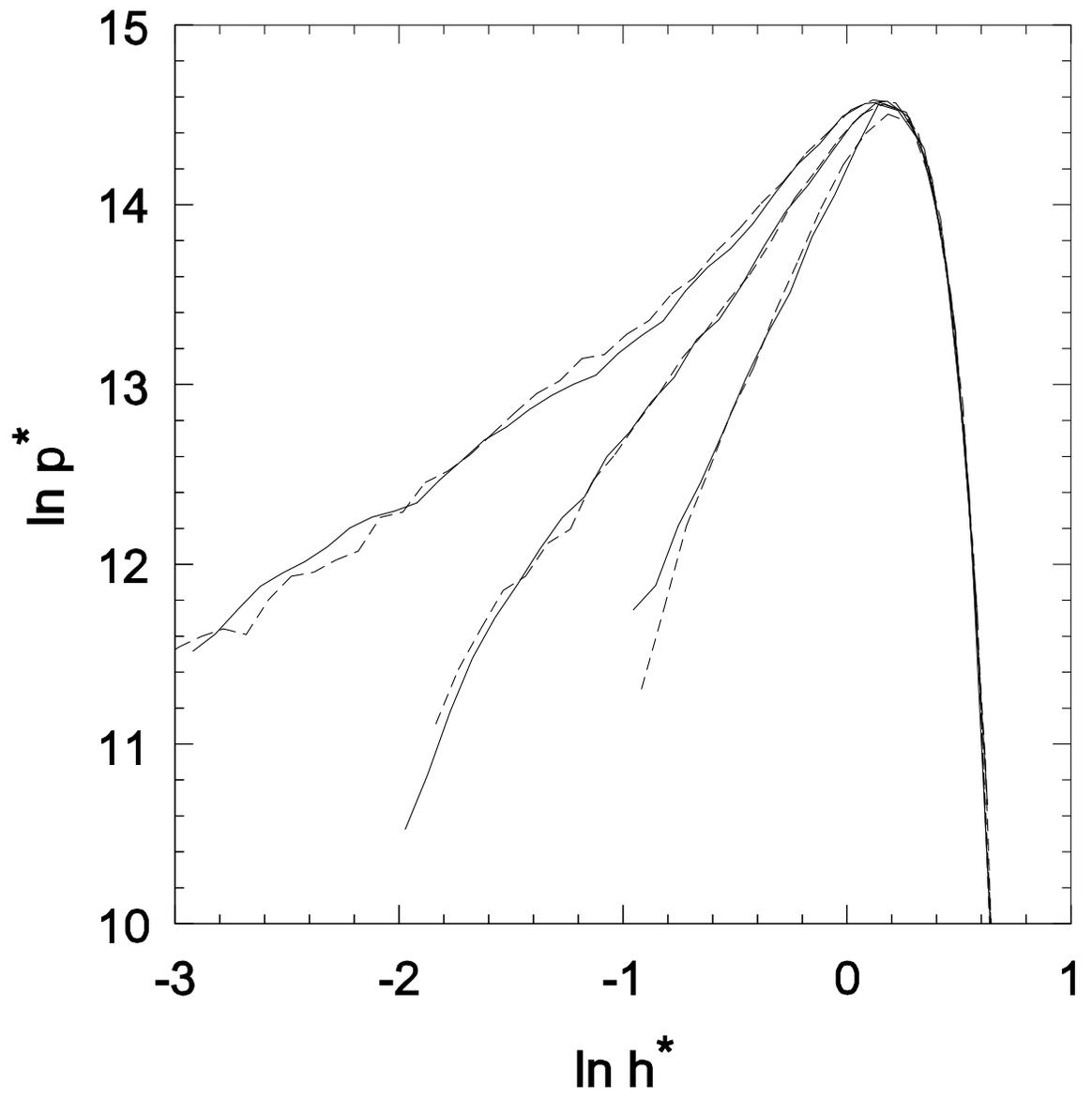
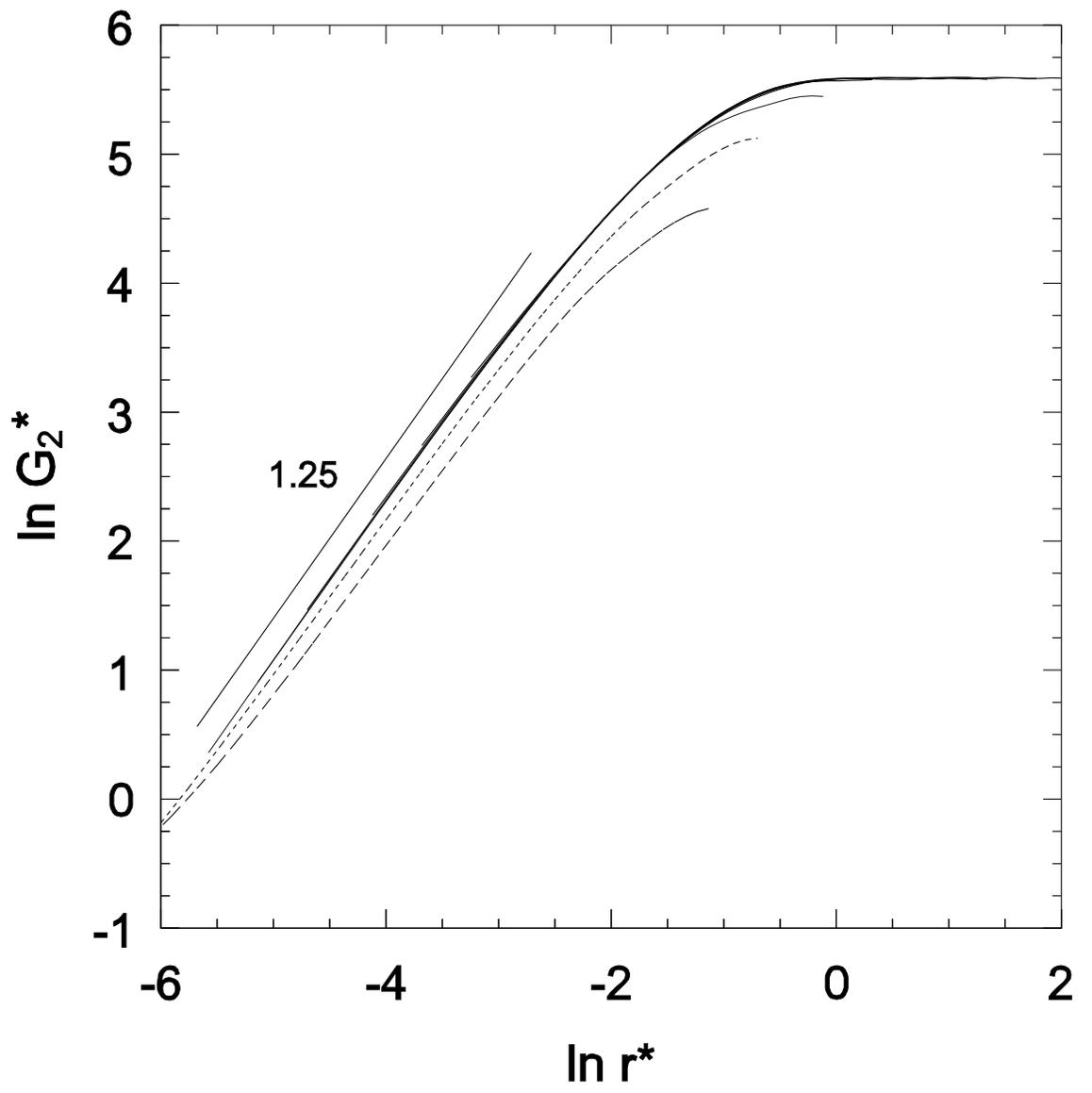
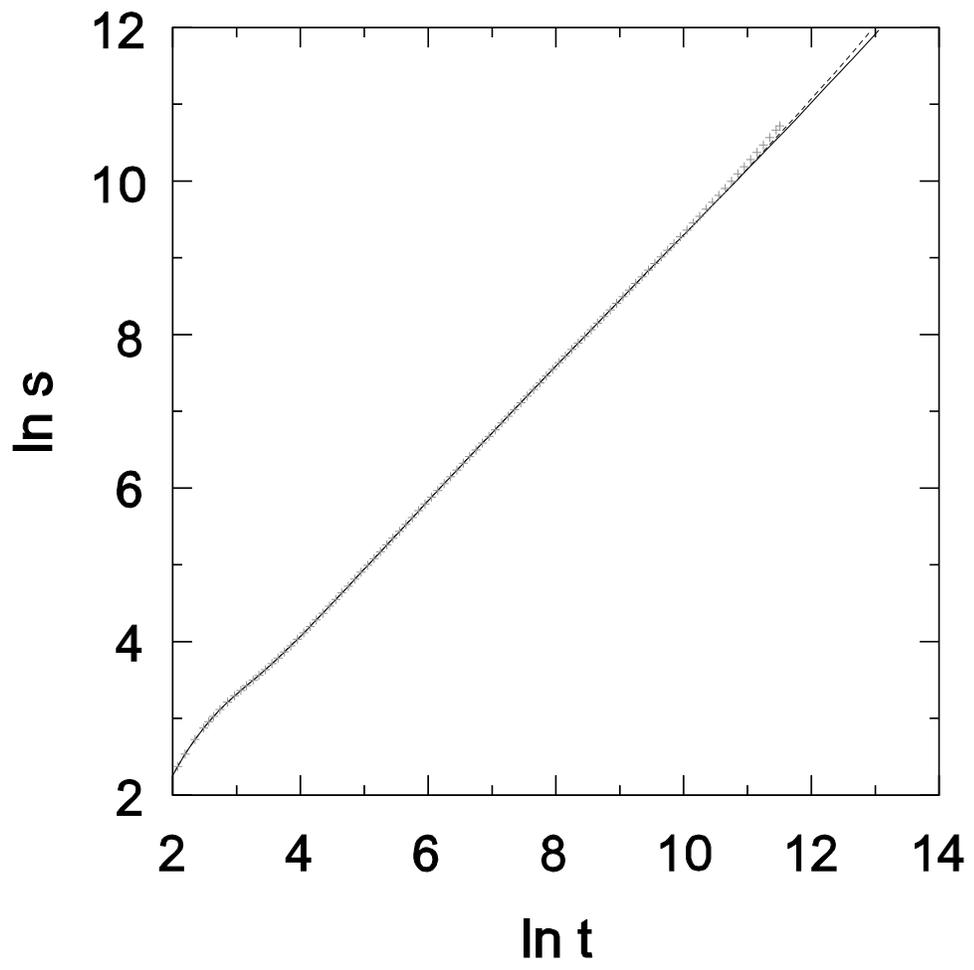
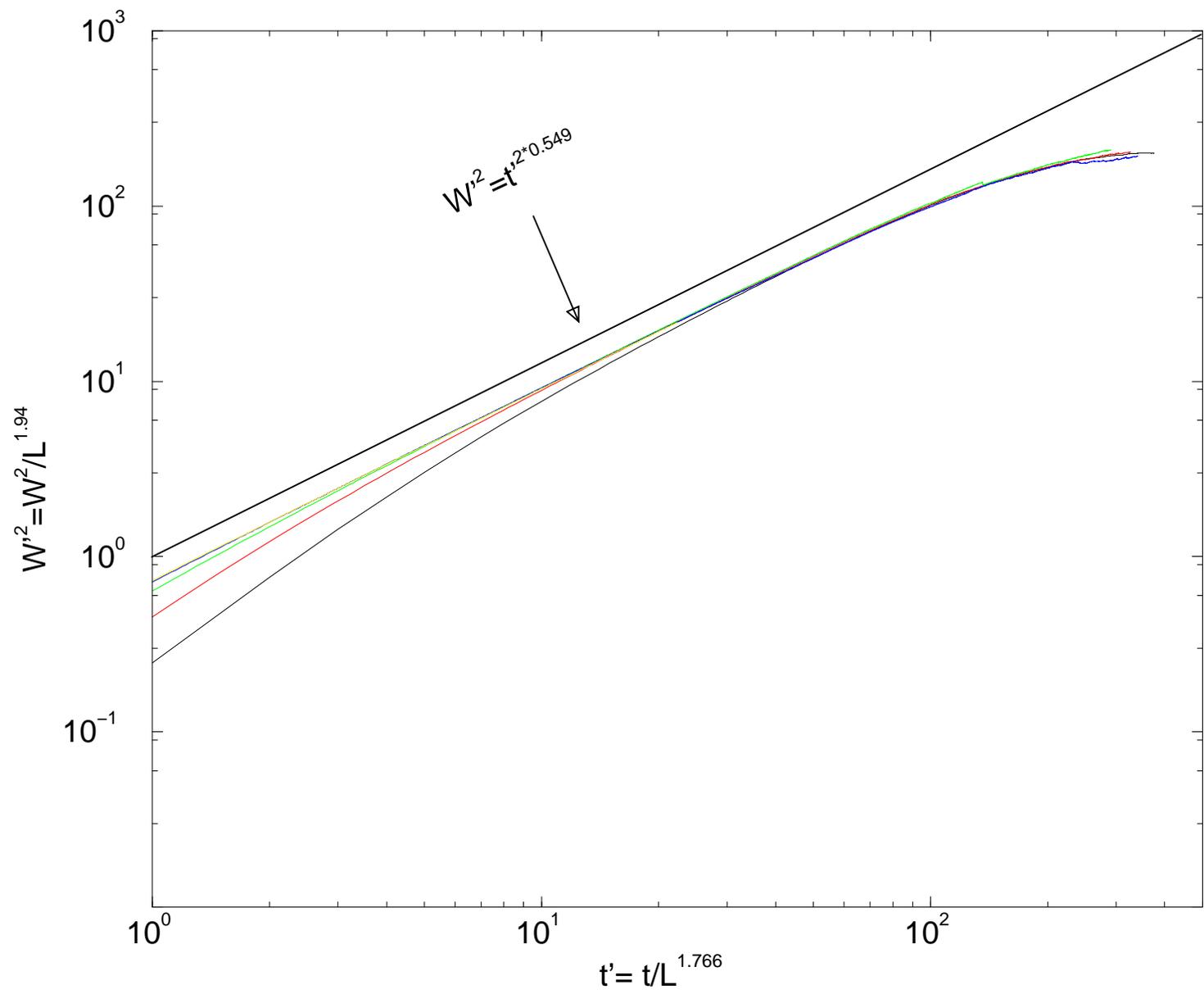
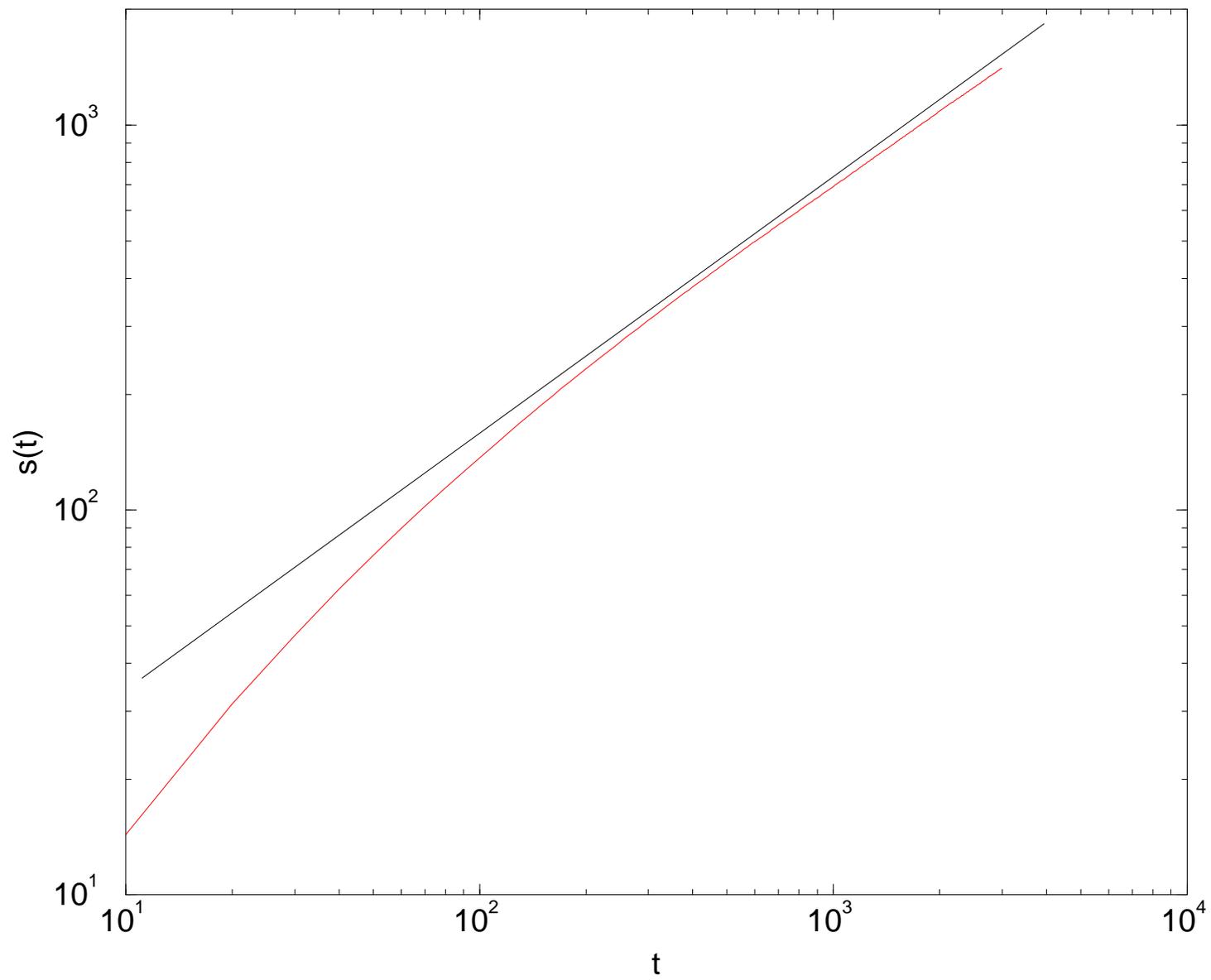
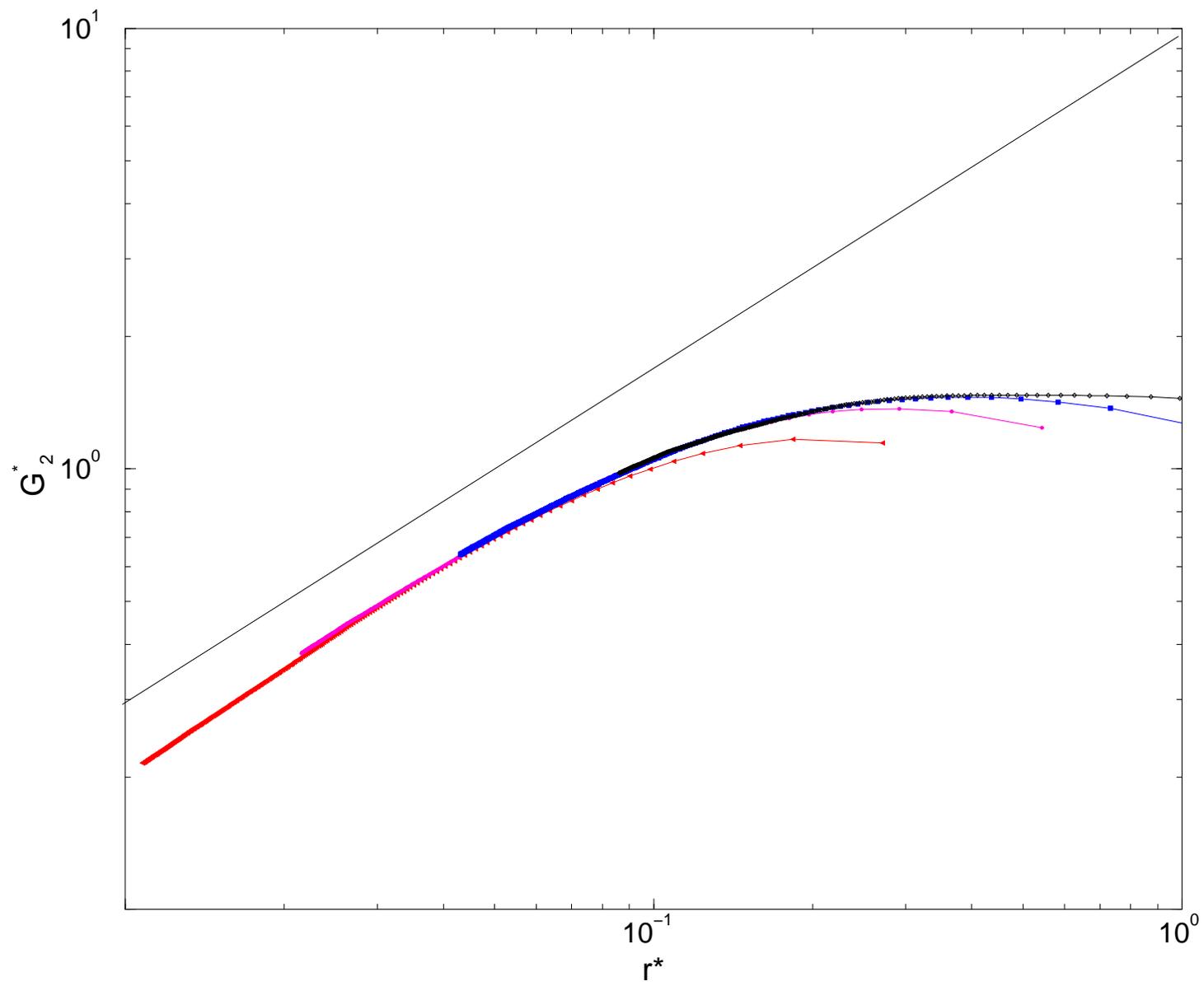
\end{document}